\documentclass[10pt,letterpaper]{article}
\usepackage[top=0.85in,left=2.35in,footskip=0.75in, marginparwidth=2in]{geometry}

\usepackage[utf8]{inputenc}

\usepackage{cite}

\usepackage{nameref,hyperref}

\usepackage[right]{lineno}
\usepackage{appendix}
\usepackage{microtype}
\DisableLigatures[f]{encoding = *, family = * }

\raggedright
\usepackage{changepage}

\usepackage[aboveskip=1pt,labelfont=bf,labelsep=period,singlelinecheck=off]{caption}

\makeatletter
\renewcommand{\@biblabel}[1]{\quad#1.}
\makeatother

\usepackage{lastpage,fancyhdr,graphicx}
\usepackage{epstopdf}
\fancyheadoffset[L]{2.25in}
\fancyfootoffset[L]{2.25in}

\usepackage{color}

\definecolor{Gray}{gray}{.25}

\usepackage{graphicx}

\usepackage{sidecap}
\usepackage{amssymb}
\usepackage{amsmath}
\usepackage{wrapfig}
\usepackage[pscoord]{eso-pic}
\usepackage[fulladjust]{marginnote}
\reversemarginpar

\begin{document}
\vspace*{0.35in}

\begin{flushleft}
{\Large
\textbf\newline{Many-levelled continuation ratio models for frequency of alcohol and drug use data.}
}
\newline
\\
Mark Chambers\textsuperscript{1,*},
Christopher Drovandi\textsuperscript{2,3}
\\
\bigskip
\bf{1} National Drug and Alcohol Research Centre, University of New South Wales, Sydney, Australia.
\\ \vspace{1.5mm}
\bf{2} School of Mathematical Sciences and Centre for Data Science, Queensland University of Technology, Brisbane, Australia. \\ \vspace{1.5mm}
\bf{3} Australian Research Council Centre of Excellence for Mathematical and Statistical Frontiers, Melbourne, Australia.
\\
\bigskip
* mark.stanley.chambers@gmail.com

\end{flushleft}

\section*{Abstract}
Studies of alcohol and drug use are often interested in the number of days that people use the substance of interest over an interval, such as 28 days before a survey date. Although count models are often used for this purpose, they are not strictly appropriate for this type of data because the response variable is bounded above. Furthermore, if some peoples’ substance use behaviors are characterized by various weekly patterns of use, summaries of substance days-of-use used over longer periods can exhibit multiple modes. These characteristics of substance days-of-use data are not easily fitted with conventional parametric model families. We propose a continuation ratio ordinal model for substance days-of-use data. Instead of grouping the set of possible response values into a small set of ordinal categories, each possible value is assigned its own category. This allows the exact numeric distribution implied by the predicted ordinal response to be recovered. We demonstrate the proposed model using survey data reporting days of alcohol use over 28-day intervals. We show the continuation ratio model is better able to capture the complexity in the drinking days dataset compared to binomial, hurdle-negative binomial and beta-binomial models.


\section*{Introduction}
Data measuring frequency of substance use are often collected with questions such as ``How many days of the last 28 did you use …''. Valid answers to this question are limited to the 29 integer values between 0 and 28. In some cases it is sufficient to partition the full set of possible values into a smaller number of ordered frequency classes and treat the outcome as ordinal \cite{Alkan2021,Janulis2021}. However, often models are fitted to the numeric responses. In this case, the choice of an appropriate response distribution warrants careful thought. 

Substance days-of-use is often modelled as count data using Poisson or, more commonly, negative binomial models\cite{Bahr2005,Lundborg2002} or their zero-inflated or hurdle variants\cite{Cranford2010,Johnson2011}. However, the number of days of substance use over a fixed interval is a bounded count since values greater than the length of the interval in days are not possible\cite{Agresti2010,Britt2018}.The set of possible observed values are the same as would be modelled by a binomial model where each observation represents the sum of 28 Bernoulli trials. However, the appropriateness of a binomial model for substance use data is also questionable because, in the application described here, the binomial assumption would require that use of the substance on each of the 28 days was independent of whether it was used on any and all of the other 27 days. If, over a given 28-day interval, a person may be consistently more or less inclined to use the substance than explained by covariates, observed substance days-of-use will exhibit more variance than expected according to a binomial distribution. Therefore, the ability of the beta-binomial model to accommodate overdispersion\cite{Wagner2015} likely makes it a better candidate than the binomial for substance days-of-use data. 

Another feature sometimes observed in days-of-use data over a 28-day period is a preponderance of observations consistent with repeated weekly patterns of use. For example, if some people regularly consumed alcohol on weekends, but rarely on other days, modes in days-of-use at four and eight days out of 28 might be observed.  Although the binomial and beta binomial models have the same support as the data, they cannot capture this multi-modality.

We propose the use of continuation ratio models \cite{Tutz1991}, a type of sequential ordinal model, for substance days-of-use data. Sequential ordinal models can be appropriate when the attainment of a particular level requires first attaining all lower levels\cite{Allison2012,Liu2019,Tutz2021}. 

The use of an ordinal model allows the distribution of the response variable to be constrained to plausible values, but also allows the odds of progressing beyond each threshold of use to differ. This second feature permits fitting distributions for substance days-of-use with multiple modes. We suggest considering a separate category for each observed value of substance days-of-use. Specifying a separate level for each observed value is less arbitrary than allocating the 29 possible response values into some smaller number of ordinal categories. More importantly, allocating a unique ordinal level for each possible response value facilitates the calculation of numeric summaries such as posterior means, quantiles and variances without needing to resort to crude approximations \cite{Aitkin2008}. 

We demonstrate the approach using reported alcohol drinking days by individuals over four 28-day intervals during the COVID-19 pandemic in Australia. We compare results obtained with the proposed continuation ratio model with equivalent models assuming hurdle-negative binomial, binomial and beta-binomial response distributions. 

\section*{Materials and Methods}

\subsection*{Data considered}
Online surveys were used to collect data on substance use of Australians during the COVID-19 pandemic for the Australians’ Drug use: Adapting to Pandemic Threats (ADAPT)  study. People were eligible to be recruited to the study if they were at least 18 years of age and had used illicit drugs regularly (i.e., at least once a month) in 2019 (i.e., `pre-COVID'). Participants had the option to complete a one-off baseline survey, or to consent to subsequent follow-up surveys (2 months, 6 months and 12 months post baseline). this paper uses data from the latter group (n=452). Participants were reimbursed \$15AUD (GiftPay voucher sent via email) for each survey that they completed, and also went in the draw to win one of three \$100 AUD GiftPay vouchers. Data were collected on use of a range of substances, price and ease of obtaining illicit substances as well as how respondents had been impacted by the pandemic at each wave of the survey. Ethical approval was granted by the UNSW Sydney Human Research Ethics Committee (HC200264). 

\subsection*{Outcome variable}
The outcome variable modelled is the self-reported number of days that alcohol was used over the 28 days prior to each survey date. We model use of alcohol data generated from responses to two questions. Abstinence from alcohol use over the 28 days was deduced from the question “When did you LAST use alcohol?”. Respondents that indicated they had never used alcohol, or they last used more than four weeks before the survey date were assigned a value of zero alcohol drinking days. Non-zero values were assigned, based on responses to the question “In the past four weeks (28 days) how many days have you drunk alcohol?”. Responses were selected from a dropdown menu with options including “1 (once)”, “2 (once a fortnight)”, “3”, “4 (once a week)”, “5”, and so on. That is, some options were described as the result of a specific pattern of drinking, while others were not. For the continuation ratio model, the response variable was defined as an ordinal type with a separate category for each of the 29 possible values from zero to 28. An integer-valued version of the same variable was used when other model families were fitted.

The data are longitudinal with surveys from 452 people initially considered in Wave 1. Responses from 436 of these participants had complete covariate data and were used to condition models. Then, 302, 258 and 250 completed surveys from the original sampled population were used from Waves 2 to 4 respectively. The decline in sample size with survey wave was mainly due to loss to follow up. Few surveys were submitted missing values of the explanatory variables. 

\subsection*{Explanatory variables}
The outcome variable was regressed on five explanatory variables. These were \texttt{survey wave}, \texttt{isolation}, \texttt{gender}, \texttt{rurality} and \texttt{state}. Survey wave was treated as a categorical variable with four levels, Wave 1 was May and June 2020, Wave 2 was July to September 2020, Wave 3 was predominantly November and December 2020, and Wave 4 was between May and July 2021. Isolation was defined as a dichotomous variable and coded as ``yes'' if the participant reported home-isolation or 14-day quarantine during the period of interest, otherwise the variable was coded as “no”. Gender was coded as a factor with 3 levels, male, female and non-binary. Rurality was a dichotomous variable, coded as either city or rural/regional. There were few respondents from any of the states of Tasmania, the Northern Territory and the Australian Capital Territory. Therefore, we combined respondents from Tasmania with Victoria, those from the Northern Territory with South Australia and those from the Australian Capital Territory with New South Wales. A person-level random intercept was specified to allow for differences among people in overall alcohol consumption after adjusting for the other covariates. 

\subsection*{Statistical analysis}

We fitted Bayesian regression models to number of alcohol drinking days over four waves of the ADAPT study during the COVID-19 pandemic. In the first instance, we assumed the probability that person $i$ consumes alcohol on $D_{ij} = 0, \dots , 28$ days during wave $j$ is described by a continuation ratio ordinal model with 29 levels. The brms package\cite{Burkner_brms_package_jss} was used to conveniently code and run Bayesian models in R version 4.1.2\cite{R_core_team} by invoking the Stan Bayesian model fitting software\cite{stan}. The goodness of fit of the ordinal posterior distribution of the continuation ratio model is compared with models assuming alternative response distributions regressed on the same explanatory variables using posterior predictive checks\cite{Gabry2021} and the leave one out information criterion (LOO-IC)\cite{Vehtari2017}.

\subsubsection*{Dependence of linear predictors on explanatory variables}
Differences among people and survey waves in number of drinking days are assumed to be explained by differences in the values of distributional parameters of the specified response distribution family. Distributional parameters are related to a linear combination of the explanatory variables via a monotonic link function as commonly applied in generalized linear mixed models (GLMMs). Let $\mathbf{x}_{ij}$ be a row vector of explanatory variables for COVID-19 isolation status, wave, state, rurality, person $i$ on wave $j$ and $\boldsymbol{\beta}$ be a column vector of coefficients. Then the linear 
predictor, $\eta_{ij}$, is specified as 

\begin{equation}
	\label{eqn:linear_pred}
\eta_{ij} = \mathbf{x}_{ij}\boldsymbol{\beta} + b_i,\quad b_i \sim \mathrm{Normal}\left(0,\sigma_b^2 \right)
\end{equation}

where $b_i$ is a person-level random intercept. Distributional parameters from each fitted model are regressed against the explanatory variables as described in Equation \ref*{eqn:linear_pred}. It should be recognized that while $\mathbf{x}_{ij}$ is the same in all models, the parameters $\boldsymbol{\beta}$ and $b_i$ differ between models and distributional parameters.

In GLMMs, the dependence of an outcome variable of interest on predictors is usually modelled explicitly through a single \textit{location} distributional parameter such as the mean of the assumed response distribution. The values of other distributional parameters are assumed to be the same for all responses. So called models for “location, scale and shape”\cite{Rigby2005} are a generalization of this approach. Similarly, the brms package allows all distributional parameters of the specified response distribution to be modelled as functions of explanatory variables. For both the hurdle-negative binomial and beta-binomials that we fitted, we specified dependence on explanatory variables for two distributional parameters. In these cases, correlation between the random effects, $b_i$, for each person, $i$, in the regression models for the two distributional parameters was modelled\cite{Burkner_jss_irt}.

\subsubsection*{Continuation ratio models with logit link}
The continuation ratio model is a sequential ordinal model in that each response is modelled as the result of a sequence of binary outcomes. We describe here, the simplest case in which, for each person, the probability of each potential progression to another day’s use in a 28-day interval is controlled by the same value, $\eta_{ij}$. A threshold parameter, $\theta_r$, quantifies a degree of restraint from progressing to at least $r$ drinking days given the person consumes alcohol on at least $r-1$ days. Initially, setting $r=1$, each person is assumed to participate in a decision to drink or not drink alcohol on at least one day. People participate in a second decision to drink on at least two days if and only if they pass the threshold of drinking on a first day, and so on until the participant reaches their total drinking days. A latent variable motivation for sequential ordinal models is described in Tutz\cite{Tutz1991}.

We assumed the default logit link function for the continuation ratio model. Therefore, letting $D_{ij}$ be the number of alcohol drinking days by person $i$ during wave $j$, the probability of progressing to at least $r$ drinking days given at least $r-1$ is given by

\begin{equation}
	\label{eqn:cont_ratio}
\mathrm{Pr}\left(D_{ij}\ge r|D_{ij}\ge r - 1,\eta_{ij},\theta_r\right) = \frac{\mathrm{exp}\left(\eta_{ij} - \theta_r\right)}{1 + \mathrm{exp}\left(\eta_{ij} - \theta_r\right) }
\end{equation}

It can be seen from Equation \ref*{eqn:cont_ratio} that under the logistic model, the threshold parameters $\theta_r$ provide multiplicative effects, $\mathrm{exp}\left(-\theta_r\right)$, on the odds of each potential progression from $r-1$ to $r$ drinking days. Although conceptualized as a sequential process, continuation ratio models can be fitted as a single logistic regression model as described by Armstrong and Sloan \cite{Armstrong1989}. However, with the brms package, the continuation ratio model is fitted with the \texttt{cratio} model family, obviating the need to restructure the fitted data. The full probability distribution of the $\operatorname{C-Ratio}\left(\eta_{ij},\theta_1, \dots ,\theta_{28} \right)$ model is given in the Appendix. The $i$ and $j$ subscripts in the preceding model notation and hereafter indicate distributional parameters that were regressed on the explanatory variables.

\subsubsection*{Other models}
Corresponding hurdle-negative binomial, binomial and beta-binomial models were also fitted to the alcohol drinking days data for comparison with the continuation ratio model. Distributional parameters of all models were regressed against the same set of explanatory variables. Probability mass functions, expressed in terms of the distributional parameters described below, are provided in the Appendix.
\paragraph{Hurdle-negative binomial}
According to the hurdle-negative binomial model, each person, $i$, has some probability,  $\psi_{ij}$, of abstaining completely from drinking alcohol on wave $j$, with $\psi_{ij}$ assumed to be described by a model similar to Equation \ref*{eqn:linear_pred} with a logit link function. The distribution of non-zero drinking days is assumed to be described by a truncated negative binomial with location parameter, $\mu_{ij}$, the mean of the (untruncated) negative binomial, is also defined similar to Equation \ref*{eqn:linear_pred}, but with a log link function. We assume the overdispersion or scale parameter, $\alpha$, of the negative binomial takes the same value for all people and waves. Normally, the hurdle component and the truncated negative binomial component are two distinct models that can be fitted separately \cite{Welsh1996}. Here, we model correlation between the person-level random effects of the $\psi_{i\boldsymbol{\cdot}}$ and $\mu_{i\boldsymbol{\cdot}}$. According to the hurdle-negative binomial model, alcohol drinking days, $D_{ij}$ are realisations of a $\operatorname{Hurdle-NB}\left(\psi_{ij}, \mu_{ij}, \alpha \right)$ distribution as defined in the Appendix.

\paragraph{Binomial}
According to the fitted binomial model, the expected probability, $\pi_{ij}$, that person $i$ drinks alcohol on a given day in wave $j$ is assumed to be defined by Equation \ref*{eqn:linear_pred} via a logit link function. Then, the number of days person $i$ consumed alcohol over 28 days during wave $j$, $D_{ij}$, is assumed to be the realization of a $\operatorname{Binomial}\left(28,\pi_{ij}\right)$ distribution.

\paragraph{Beta-binomial}
The beta-binomial model is parameterized in terms of an underlying (binomial) probability of drinking, $\pi_{ij}$, and an overdispersion parameter, $\phi_{ij}$. The dependence of $\pi_{ij}$ and $\phi_{ij}$ on explanatory variables is parameterized as described in Equation \ref*{eqn:linear_pred} with logit and log links respectively, except intra-person correlation between the random effects for each distributional parameter is modelled. The beta-binomial distribution is not included among native (inbuilt) brms model families. However, package documentation gives instructions on how it can be user-defined\cite{Burkner_webpage}. We denote the fitted beta-binomial model $\operatorname{Beta-Bin}\left(28,\pi_{ij},\phi_{ij}\right)$ where the (beta-binomial) probability that person $i$ drinks alcohol on a given day in wave alcohol on a given day in wave $j$, $\pi_{ij}^*$, is a random variable with a beta distribution, parameterized in terms of $\pi_{ij}$ and $\phi_{ij}$ (see Appendix).

\subsubsection*{Prior distributions and Monte Carlo details}
We used horseshoe priors\cite{Carvalho2009} for population-level (i.e.\,fixed effects) coefficient parameters to provide a level of protection against overfitting. For other parameters we used brms default priors. These are $t$ distributions with three degrees of freedom and a scale parameter of 2.5 for both the ordinal level thresholds, $\theta_r$, and for the standard deviations of the random effects distributions, $\sigma_b$. For models with multiple distributional parameters, the Lewandowski-Kurowicka-Joe (LKJ) prior distribution\cite{Lewandowski2009} was assumed for the random effect correlation matrices. The LKJ prior is the brms default in this case.

Posterior distributions of model parameters were approximated with four chains of 5000 iterations each generated by Stan’s No U-turn Sampler. The first 500 iterations from each chain were discarded as burn-in or warm-up after which every fifth iteration was kept, resulting in a total of 3,600 samples from each conditioned model on which to base posterior inference.  Convergence of the Monte Carlo algorithms was assessed by visual inspection of mixing of the four chains in trace plots and the rank-based $\hat{R}$ statistic convergence diagnostic\cite{Vehtari2021}.

\section*{Results}
\subsection*{Distribution of reported drinking days}
The alcohol drinking days data exhibits periodic modes every four days, consistent with repeated drinking patterns of once-a-week, twice-a-week, and so on (Fig. \ref{fig1}). Zero alcohol use was reported on approximately 17 percent of 28-day intervals across the four waves combined.
\subsection*{Posterior predictive checks}
\marginpar{
	\vspace{1.5cm} 
	\color{Gray} 
	\textbf{Figure 1. Alcohol drinking days.} 
	Reported days of alcohol use during 28-day intervals across four survey waves of the ADAPT study. Note the logarithmic y-scale.
}
\begin{wrapfigure}[16]{l}{75mm}
	\includegraphics[width=75mm]{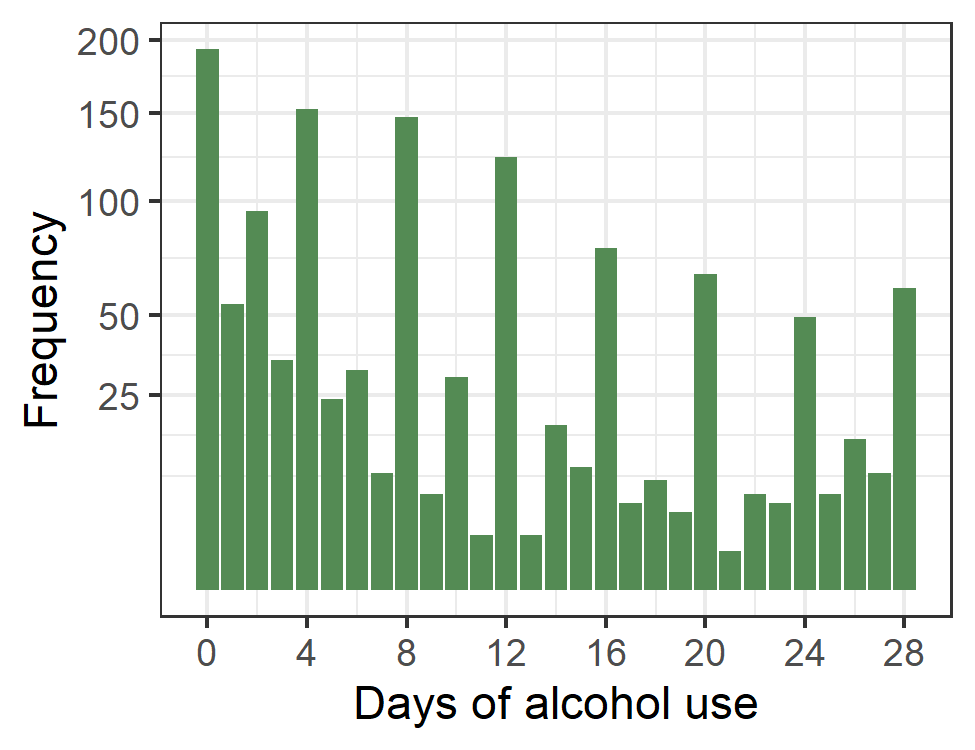}
	\captionsetup{labelformat=empty} 
	\caption{} 
	\label{fig1} 
\end{wrapfigure}  

\begin{figure}[ht] 
	\includegraphics[width = \textwidth]{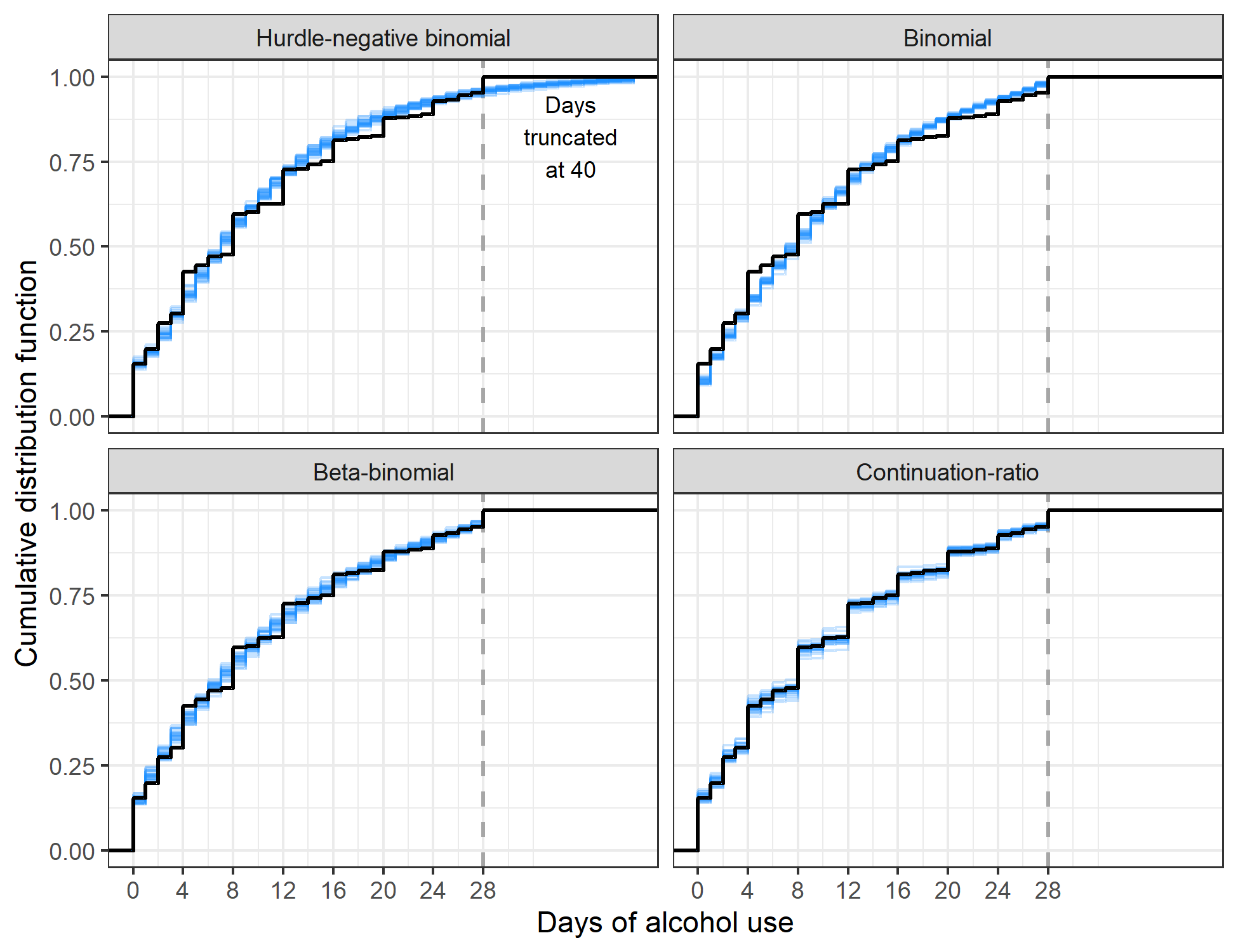}
	
	\caption{\color{Gray} \textbf{Posterior predictive ECDF check}. Posterior predictive checks based on the empirical cumulative distribution function (ECDF) of days of alcohol used across all four waves of the ADAPT study. Observed ECDFs of days of alcohol use over four 28-day intervals in Australia during the COVID-19 pandemic are shown with black lines. The thin blue lines are ECDFs based on 25 draws from the posterior predictive distribution of the hurdle-negative binomial, binomial, beta-binomial and continuation ratio models conditioned on the same data. Predictions from the hurdle-negative binomial model have been truncated at 40 days for convenience of presentation.}
	
	\label{fig2} 
\end{figure}

The observed empirical cumulative distribution function (ECDF) for drinking days across the four waves is depicted by the (identical) black line  in each panel of Figure \ref{fig2}. The multi-modal distribution of reported drinking days, evident in Figure \ref{fig1}, manifests as large steps in the observed ECDF at multiples of four days, relative to smaller steps in the ECDF observed at other days of alcohol use (Fig. \ref{fig2}). The thin blue lines in Figure \ref{fig2} are ECDFs resulting from 25 draws from the posterior predictive distribution (see Gelman et al.\cite{Gelman2013} Chapter 6) of each model. 

\begin{figure}[ht] 
	\includegraphics[width = \textwidth]{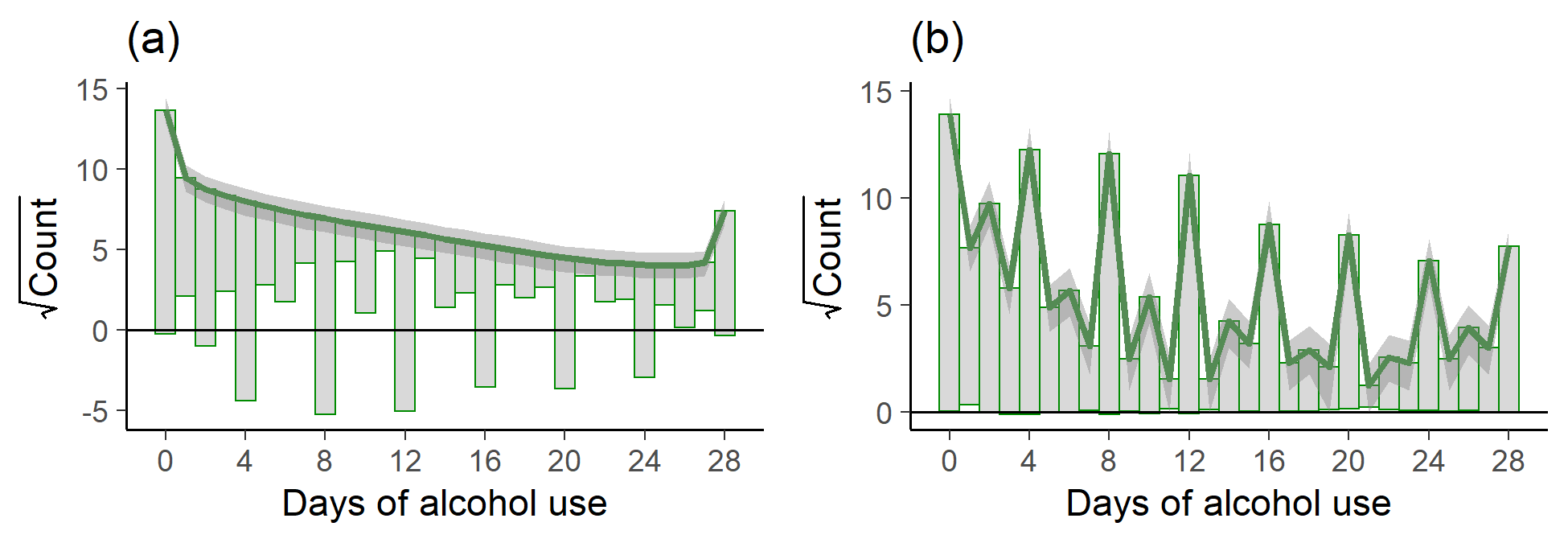}
	
	\caption{\color{Gray} \textbf{Posterior predictive rootogram check}. Posterior predictive hanging rootogram check for (a) beta-binomial and (b) continuation ratio models. Green lines are posterior predictive medians for the square roots of the predicted frequencies, the hanging grey bars have heights equal to the square root of the observed frequencies of each number of drinking days. The difference between the bottoms of the bars and the zero line indicates the square roots of the difference between the posterior predicted medians and observed values.}
	
	\label{fig3} 
\end{figure}

Comparing the observed and posterior predictive ECDFs for each model indicates the binomial model predicts both fewer responses near zero and fewer responses near 28 than are reported. This suggests that the reported drinking days are overdispersed compared with what would be expected if drinking days had a binomial distribution. A problem with the hurdle-negative binomial model is that a small number of the posterior predicted values exceed 28 days use on each MCMC iteration. Predictions from the hurdle-negative binomial model for Figure \ref{fig2} were truncated at 40 drinking days for ease of presentation.  The posterior predictive distribution of the beta-binomial model reproduces the coarse scale distribution in reported drinking days but is unable to capture the heterogeneity in adjacent step sizes in the ECDF evident in observed drinking days. In contrast to the other models, the continuation ratio model closely reproduces the heterogeneity in step size in the ECDF. The difference in the posterior predictive distributions of the beta-binomial and ordinal models is more clearly shown in rootograms (Fig. \ref{fig3}).
	
\subsection*{Numeric summaries}

\begin{figure}[ht] 
	\includegraphics[width = \textwidth]{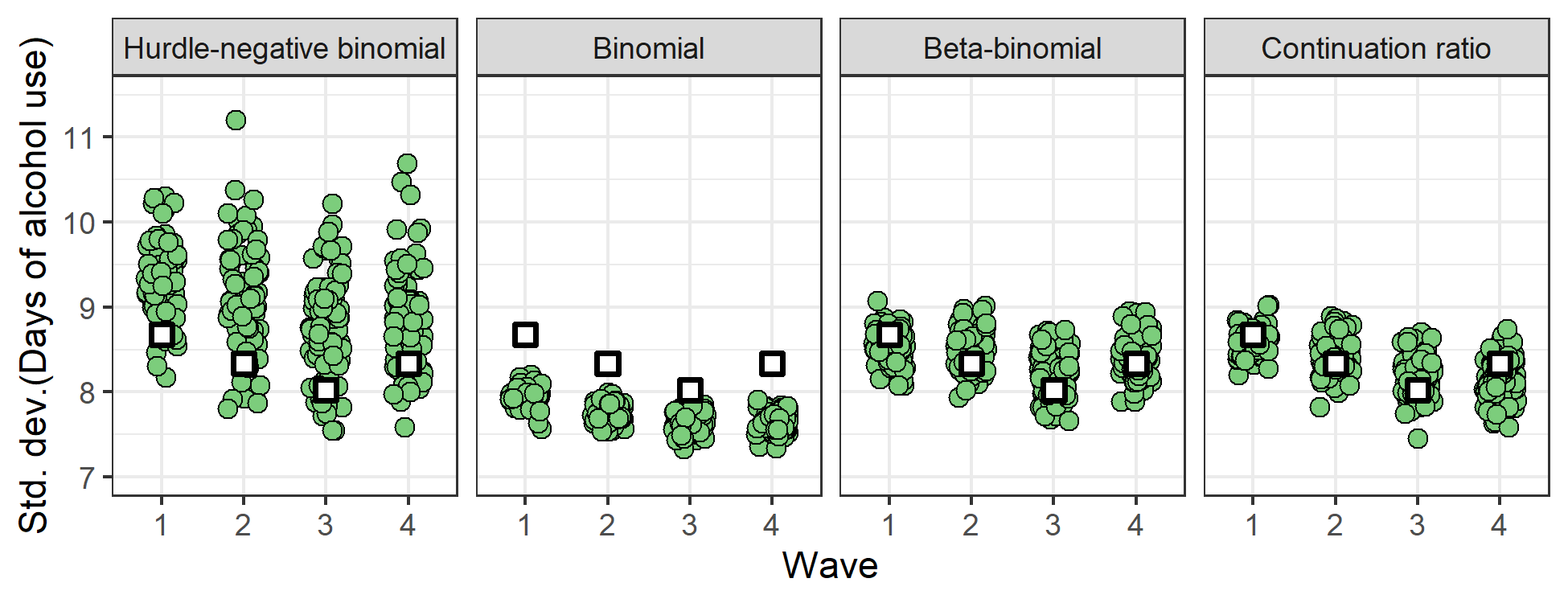}
	
	\caption{\color{Gray} \textbf{Posterior predictive standard deviation check}. Wave-specific standard deviations of posterior predicted drinking days for the fitted survey participants according to 320 draws from the posterior distributions of the four conditioned models. The observed wave-specific standard deviations are depicted by the open squares.}
	
	\label{fig4} 
\end{figure}

Although the continuation ratio model treats the response as ordinal, the full implied numeric distribution can be recovered from its posterior distribution. All four models predicted mean alcohol drinking days by survey wave reasonably well (not shown). However, there were differences in posterior predictive standard deviation (Fig. \ref{fig4}). The standard deviation of the hurdle-negative binomial posterior predictive distribution was more variable than the other models. The binomial posterior predictive distribution has lower standard deviation than the observed data, further highlighting overdispersion in the observed data relative to the binomial model assumption. The beta-binomial and continuation ratio models both give posterior predictions for each survey wave with standard deviations consistent with the observed data.

\begin{figure}[ht] 
	\includegraphics[width = \textwidth]{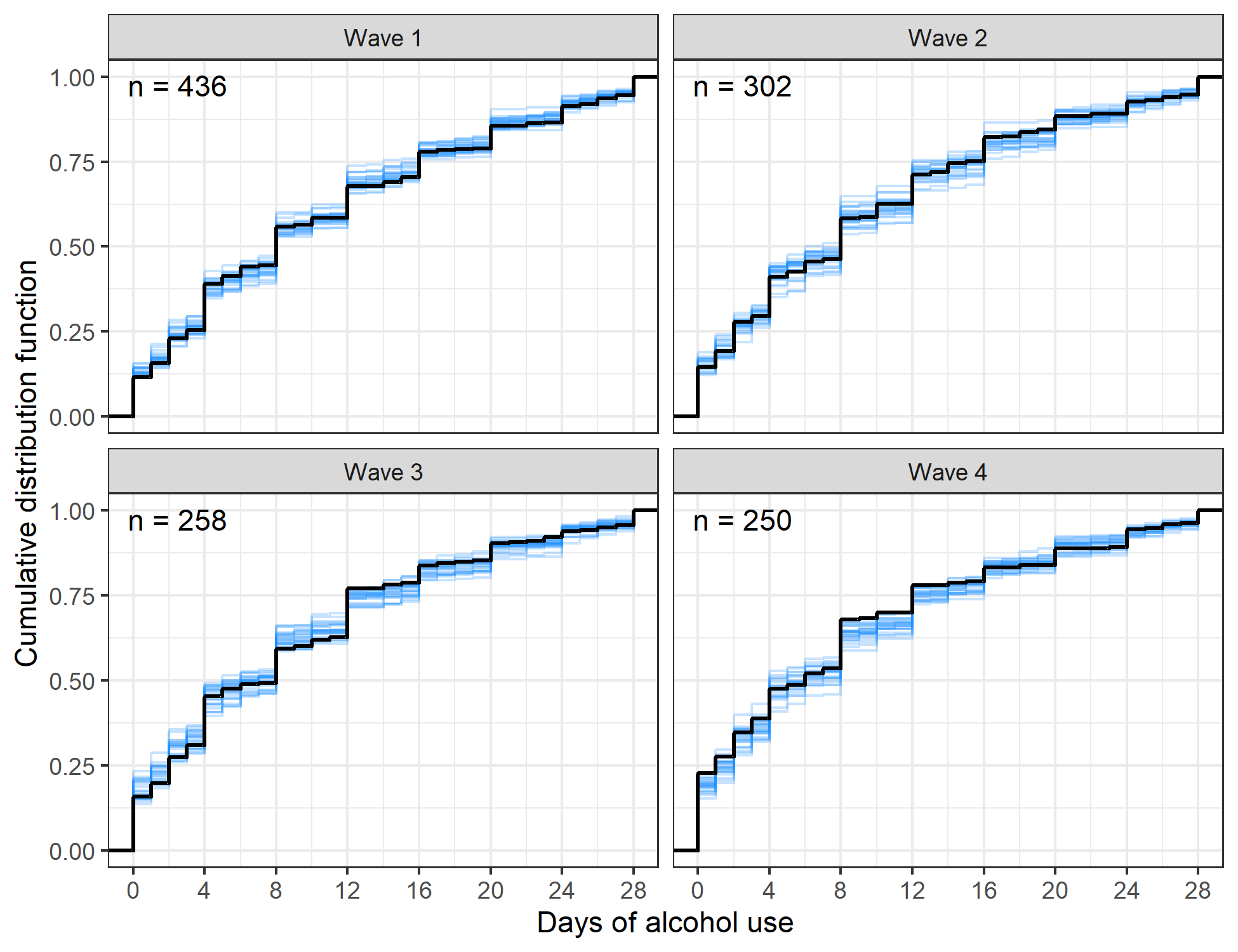}
	
	\caption{\color{Gray} \textbf{Posterior predictive standard deviation check}.Posterior predictive check of empirical cumulative distribution functions predicted by the continuation ratio ordinal model disaggregated by survey wave. Thin blue lines are ECDFs predicted from 25 random draws from the posterior distribution.}
	
	\label{fig5} 
\end{figure}

It would be expected that a sequential model with 28 thresholds would fit cumulative days use well overall (Fig. \ref{fig2}). A more rigorous test of this model is to check whether the posterior distribution of the thresholds for progression of alcohol drinking days are reasonable for different values of explanatory variables. Separate ECDFs for the four survey waves (Fig. \ref{fig5}) reveal complete abstinence from drinking over the 28-day intervals increased from 14\% in Wave 1 to 24\% in Wave 4. The posterior predicted distribution of the ECDF for each wave is reasonably consistent with the observed data.

\subsection*{LOO-IC}
Although the same set of explanatory variables was used for each model, the four models do not have the same number of parameters. For example, the continuation ratio model includes the 28 threshold parameters that are not included in the other models. On the other hand, the hurdle-negative binomial and the beta-binomial have two sub-models with full sets of parameters, including person-level random effects, estimated for each. Since the continuation ratio model includes a separate ordinal category for each numeric response, we can compare its goodness of fit with the other three models using information criteria. According to LOO-IC (Table \ref{table:model_loos}), the continuation ratio model fits the alcohol drinking days data clearly better than the other models, accounting for differences in the number of parameters between models.

\begin{table}[!ht]
	\caption{Leave one out information criterion (LOO-IC) for fits of the four models to the alcohol drinking days data with estimated standard errors. P-LOO is the estimated number of effective parameters in each model. Lower values of LOO-IC indicate better fit.}
	\begin{minipage}{14 cm}
		\centering
		\begin{tabular}{l r r}
					\hline
		\rule{0pt}{2.5ex}	\textbf{Model}  & \textbf{P-LOO (SE)} &  \textbf{LOO-IC (SE)} \\ \hline
		\rule{0pt}{2.5ex}	$\operatorname{C-Ratio}\left(\eta_{ij},\theta_1,\cdots , \theta_{28}\right)$ & 357 (13.8) & 5665 (89.1)\\
		\rule{0pt}{2.5ex}	$\operatorname{Beta-Bin}\left(28,\pi_{ij},\phi_{ij}\right)$	& 460 (12.8) & 6473 (78.5) \\
		\rule{0pt}{2.5ex}	$\operatorname{Hurdle-NB}\left(\psi_{ij}, \mu_{ij}, \alpha \right)$ & 417 (15.8) & 6963 (86.4) \\
	\rule{0pt}{2.5ex}	$\operatorname{Binomial}\left(28, \pi_{ij} \right)$ & 1078 (51.7) & 8571 (253.2) \\
			\hline
		\end{tabular}
	\end{minipage}
	\label{table:model_loos}
\end{table}

\subsection*{Inference}
\begin{figure}[ht] 
	\includegraphics[width = \textwidth]{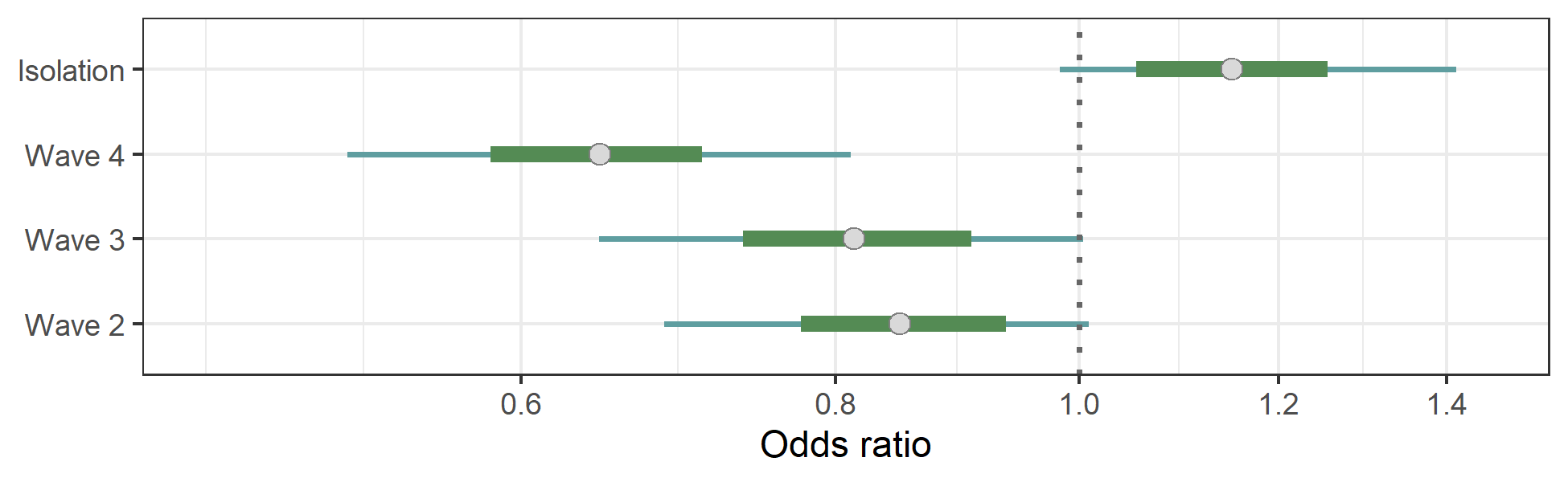}
	
	\caption{\color{Gray} \textbf{Posterior distributions of selected odds ratios}. Posterior distributions of odds ratios of extending drinking days in Waves 2, 3 and 4, relative to Wave 1 and when in isolation or quarantine relative to when not. Plotted points are posterior medians, thick green bands are 50\% credible intervals and thin gray lines are 90\% credible intervals.}
	
	\label{fig6} 
\end{figure}

In applications such as modelling substance use, it will usually be advantageous if clear inferences can be made based on the fitted model. As the best model, we base posterior inference on the continuation ratio model. Exponentiating the posterior distribution of the parameters, $\boldsymbol{\beta}$, from Equation \ref{eqn:linear_pred} gives posterior odds ratios for extending the number of alcohol drinking days levels of categorical variables relative to specified reference categories. For the ADAPT study, primary interest lay in frequency of use of alcohol and other drugs at the four survey waves as well as the effect of home isolation and quarantine. The posterior distribution of the odds ratio for isolation versus no isolation was mostly greater than unity, suggesting the target population was probably somewhat more inclined to extend their alcohol drinking days when in isolation or quarantine than when not. Based on the posterior median estimate, the odds of extending drinking days from a given level of drinking in isolation or quarantine was around 15\% higher than when not in isolation [Posterior median odds ratios 1.15, 90\% credible interval (CI) 0.99, 1.40]. Conversely, we infer that the odds of extending alcohol drinking days most likely declined after Wave 1. Most notably, odds of extending alcohol drinking days in Wave 4 is estimated to have been around 35\% lower than Wave 1 (Posterior median odds ratio 0.65, 90\% CI 0.51, 0.81) (Fig. \ref{fig6}).

\section*{Discussion}
We have demonstrated the suitability of the continuation ratio model for fitting substance days-of-use data and its superiority compared with the negative binomial, binomial and beta-binomial distributions for the fitted alcohol drinking days data. The beta-binomial and hurdle-negative binomial can each be fitted with alternative numbers of distributional parameters regressed on explanatory variables from those included in Table 1. We also fitted a beta-binomial with one distributional parameter, $\operatorname{Beta-Bin}\left(28,\pi_{ij},\phi\right)$ and hurdle-negative binomials with one and three regressed distributional parameters, $\operatorname{Hurdle-NB}\left(\psi, \mu_{ij}, \alpha \right)$ and $\operatorname{Hurdle-NB}\left(\psi_{ij}, \mu_{ij}, \alpha_{ij} \right)$. None of these alternative models were competitive with the continuation ratio model according to the LOO-IC measure.

The continuation ratio model fitted to alcohol drinking days specified explanatory variable dependence through a single distributional parameter. Provided the fitted model adequately describes the data generating process, variable dependence on a single distributional parameter greatly simplifies inference. For example, Allison\cite{Allison2012} expresses a preference for negative binomial over zero-inflated Poisson and zero-inflated negative binomial models for count data because zero-inflated components make the latter more difficult to interpret (see also\cite{Saei1996}). Although odds ratios from sequential ordinal models need to be interpreted slightly differently from most logistic models, including cumulative link ordinal models, they do have a clear interpretation in the context of substance days-of-use data.

The advantage of ordinal models will be greatest when data are multimodal. It seems likely that the wording of available responses on the questionnaire in the ADAPT study exaggerated the true extent of multimodality in alcohol drinking days by prompting respondents to preferentially choose responses corresponding to multiples of four. Nevertheless, less pronounced multimodality of the kind observed is plausible. Cannabis use data modelled by Wagner et al.\cite{Wagner2015} (their Figure 2a) appear to exhibit multiple modes. Unfortunately, Wagner et al.\, model total substance days-of-use combined over several 28-day intervals for each person, potentially obscuring multimodality in their data. Also, Kowal \& Wu\cite{Kowal2021} describe characteristics of mental health stress observations similar to substance days-of-use data considered here.  Many-levelled continuation ratio models could be used in other applications where outcomes are measured in number of days observed over a fixed interval, such as migraines, interrupted sleep, and exercise\cite{Diener2007,Litt2002,Liu2013}.




\section*{Acknowledgments}
We thank the people that participated in the ADAPT survey. Rachel Sutherland advised on the operational details of the study. CD acknowledges support from an Australian Research Council Future Fellowship (FT210100260).

\nolinenumbers

\bibliography{my_library}

\begin{thebibliography}{10}

\bibitem{Agresti2010}
A.~Agresti.
\newblock {\em Analysis of Ordinal Categorical Data}.
\newblock John Wiley \& Sons, 2 edition, 2010.

\bibitem{Aitkin2008}
M.~Aitkin.
\newblock Applications of the {Bayesian} bootstrap in finite population
  inference.
\newblock {\em Journal of Official Statistics}, 24:21--51, 2008.

\bibitem{Allison2012}
P.~D. Allison.
\newblock {\em Logistic regression using SAS: Theory and application}.
\newblock SAS institute, 2012.

\bibitem{Armstrong1989}
B.~G. Armstrong and M.~Sloan.
\newblock Ordinal regression models for epidemiologic data.
\newblock {\em American Journal of Epidemiology}, 129:191--204, 1989.

\bibitem{Bahr2005}
S.~J. Bahr, J.~P. Hoffmann, and X.~Yang.
\newblock Parental and peer influences on the risk of adolescent drug use.
\newblock {\em Journal of Primary Prevention}, 26:529--551, 11 2005.

\bibitem{Britt2018}
C.~L. Britt, M.~Rocque, and G.~M. Zimmerman.
\newblock The analysis of bounded count data in criminology.
\newblock {\em Journal of Quantitative Criminology}, 34:591--607, 6 2018.

\bibitem{Burkner_webpage}
P.~C. Bürkner.
\newblock Define custom response distributions with brms, 4 2022.

\bibitem{Burkner_brms_package_jss}
P.~C. B{\"u}rkner.
\newblock brms: An {R} package for {Bayesian} multilevel models using {Stan}.
\newblock {\em Journal of Statistical Software}, 80:1--28, 2017.

\bibitem{Burkner_jss_irt}
P.~C. B{\"u}rkner.
\newblock Bayesian item response modeling in {R} with brms and {Stan}.
\newblock {\em Journal of Statistical Software}, 100, 2021.

\bibitem{Carvalho2009}
C.~M. Carvalho, N.~G. Polson, and J.~G. Scott.
\newblock Handling sparsity via the horseshoe.
\newblock pages 73--80, 2009.

\bibitem{Cranford2010}
J.~A. Cranford, R.~A. Zucker, J.~M. Jester, L.~I. Puttler, and H.~E.
  Fitzgerald.
\newblock Parental alcohol involvement and adolescent alcohol expectancies
  predict alcohol involvement in male adolescents.
\newblock {\em Psychology of Addictive Behaviors}, 24:386--396, 9 2010.

\bibitem{Diener2007}
H.~C. Diener, G.~Bussone, J.~C.~V. Oene, M.~Lahaye, S.~Schwalen, and P.~J.
  Goadsby.
\newblock Topiramate reduces headache days in chronic migraine: A randomized,
  double-blind, placebo-controlled study.
\newblock {\em Cephalalgia}, 27:814--823, 7 2007.

\bibitem{Gabry2021}
J.~Gabry and T.~Mahr.
\newblock {bayesplot: Plotting for Bayesian Models}, 2021.
\newblock R package version 1.8.1.

\bibitem{Gelman2013}
A.~Gelman, J.~B. Carlin, H.~S. Stern, D.~B. Dunson, A.~Vehtari, and D.~B.
  Rubin.
\newblock {\em Bayesian Data Analysis}.
\newblock Chapman and Hall/CRC, 3 edition, 2013.

\bibitem{Janulis2021}
P.~Janulis, M.~E. Newcomb, and B.~Mustanski.
\newblock Decrease in prevalence but increase in frequency of non-marijuana
  drug use following the onset of the {COVID-19} pandemic in a large cohort of
  young men who have sex with men and young transgender women.
\newblock {\em Drug and Alcohol Dependence}, 223, 6 2021.

\bibitem{Johnson2011}
J.~E. Johnson, C.~C. O'Leary, C.~W. Striley, A.~B. Abdallah, S.~Bradford, and
  L.~B. Cottler.
\newblock Effects of major depression on crack use and arrests among women in
  drug court.
\newblock {\em Addiction}, 106:1279--1286, 7 2011.

\bibitem{Kowal2021}
D.~R. Kowal and B.~Wu.
\newblock Semiparametric count data regression for self-reported mental health.
\newblock {\em Biometrics}, 2021.

\bibitem{Lewandowski2009}
D.~Lewandowski, D.~Kurowicka, and H.~Joe.
\newblock Generating random correlation matrices based on vines and extended
  onion method.
\newblock {\em Journal of Multivariate Analysis}, 100:1989--2001, 10 2009.

\bibitem{Litt2002}
M.~D. Litt, A.~Kleppinger, and J.~O. Judge.
\newblock Initiation and maintenance of exercise behavior in older women:
  predictors from the social learning model.
\newblock {\em Journal of Behavioral Medicine}, 25:83--97, 2002.

\bibitem{Liu2019}
X.~Liu and H.~Bai.
\newblock Forward and backward continuation ratio models for ordinal response
  variables.
\newblock {\em Journal of Modern Applied Statistical Methods}, 18:2--16, 2019.

\bibitem{Liu2013}
Y.~Liu, J.~B. Croft, A.~G. Wheaton, G.~S. Perry, D.~P. Chapman, T.~W. Strine,
  L.~R. McKnight-Eily, and L.~Presley-Cantrell.
\newblock Association between perceived insufficient sleep, frequent mental
  distress, obesity and chronic diseases among us adults, 2009 behavioral risk
  factor surveillance system.
\newblock {\em BMC public health}, 13:84, 2013.

\bibitem{Lundborg2002}
P.~Lundborg.
\newblock Young people and alcohol: an econometric analysis.
\newblock {\em Addiction}, 97:1573--1582, 2002.

\bibitem{Alkan2021}
Ömer Alkan and E.~Güney.
\newblock Investigation of factors that affect the frequency of alcohol use of
  employees in {Turkey}.
\newblock {\em Journal of Substance Use}, 26:468--474, 2021.

\bibitem{R_core_team}
{R Core Team}.
\newblock {R: A Language and Environment for Statistical Computing}, 2021.

\bibitem{Rigby2005}
R.~A. Rigby and D.~M. Stasinopoulos.
\newblock Generalized additive models for location, scale and shape.
\newblock {\em Applied Statistics}, 54:507--554, 2005.

\bibitem{Saei1996}
A.~Saei, J.~Ward, and C.~A. McGilchrist.
\newblock Threshold models in a methadone programme evaluation.
\newblock {\em Statistics in Medicine}, 15:2253--2260, 10 1996.

\bibitem{stan}
{Stan Development Team}.
\newblock {Stan Modeling Language Users Guide and Reference Manual}, 2022.

\bibitem{Tutz1991}
G.~Tutz.
\newblock Sequential models in categorical regression.
\newblock {\em Computational Statistics \& Data Analysis}, 11:275--295, 1991.

\bibitem{Tutz2021}
G.~Tutz.
\newblock Ordinal regression: A review and a taxonomy of models.
\newblock {\em Wiley Interdisciplinary Reviews: Computational Statistics},
  2021.

\bibitem{Vehtari2017}
A.~Vehtari, A.~Gelman, and J.~Gabry.
\newblock Practical {Bayesian} model evaluation using leave-one-out
  cross-validation and {WAIC}.
\newblock {\em Statistics and Computing}, 27:1413--1432, 9 2017.

\bibitem{Vehtari2021}
A.~Vehtari, A.~Gelman, D.~Simpson, B.~Carpenter, and P.~C. Burkner.
\newblock Rank-normalization, folding, and localization: An improved (formula
  presented) for assessing convergence of mcmc (with discussion)*†.
\newblock {\em Bayesian Analysis}, 16:667--718, 2021.

\bibitem{Wagner2015}
B.~Wagner, P.~Riggs, and S.~Mikulich-Gilbertson.
\newblock The importance of distribution-choice in modeling substance use data:
  A comparison of negative binomial, beta binomial, and zero-inflated
  distributions.
\newblock {\em American Journal of Drug and Alcohol Abuse}, 41:489--497, 11
  2015.

\bibitem{Welsh1996}
A.~H. Welsh, R.~B. Cunningham, C.~F. Donnelly, and D.~B. Lindenmayer.
\newblock Modelling the abundance of rare species: statistical models for
  counts with extra zeros.
\newblock {\em Ecological Modelling}, 88:297--308, 1996.

\end{thebibliography}

\bibliographystyle{abbrv}

\begin{appendices}
	\setcounter{table}{0}
\renewcommand{\thetable}{A\arabic{table}}
\section{Probability mass functions}\label{sec:Appendix}

\begin{table}[!ht]
	\begin{adjustwidth}{-1.75in}{1in}
		\centering
		\caption{{\bf } Probability mass functions of fitted models with parameters as described in the text. The quantity $N$ in the continuation ratio, binomial and beta-binomial models is the length of the interval in days, which was 28 in the case of the ADAPT data fitted.}
		\begin{tabular}{l l}
			\hline
			\rule{0pt}{2.5ex}	\bf Model & \bf Probability Mass Function\\ \hline
				& \\
			\rule{0pt}{2.5ex}	$\operatorname{C-Ratio}\left(\eta_{ij},\theta_1,\cdots , \theta_{N}\right)$ & 	 $\operatorname{Pr}\left( D_{ij} = d\right) = \left\{ \begin{array}{ll}
				1 - \frac{\operatorname{exp}\left(\eta_{ij} - \theta_1\right)}{1 + \operatorname{exp}\left(\eta_{ij} - \theta_1\right)} & \qquad d = 0   \\ 
			
				1 - \frac{\operatorname{exp}\left(\eta_{ij} - \theta_{d+1}\right)}{1 + \operatorname{exp}\left(\eta_{ij} - \theta_{d+1}\right)}	\prod\limits_{r=1}^{d} \frac{\operatorname{exp}\left(\eta_{ij} - \theta_r\right)}{1 + \operatorname{exp}\left(\eta_{ij} - \theta_r\right)} & \qquad d = 1, 2, \dots , N-1 \\
				\prod\limits_{r=1}^{N} \frac{\operatorname{exp}\left(\eta_{ij} - \theta_r\right)}{1 + \operatorname{exp}\left(\eta_{ij} - \theta_r\right)} & \qquad d = N\end{array}\right. $ \\ 
			& \\
			$\operatorname{Hurdle-NB}\left(\psi_{ij}, \mu_{ij}, \alpha \right)$ & $\operatorname{Pr}\left( D_{ij} = d\right) = \left\{ \begin{array}{ll} 
				\psi_{ij} &\, d = 0  \\
				\frac{1 - \psi_{ij}}{\left(1 + \alpha \mu_{ij}\right)^{\frac{1}{\alpha}}} \frac{\Gamma\left(d + \frac{1}{\alpha}\right)}{\Gamma\left(\frac{1}{\alpha} \right)d!} \left(\frac{1}{1+ \alpha \mu_{ij}}\right)^{\frac{1}{\alpha}}  \left(\frac{\alpha \mu_{ij}}{1+ \alpha \mu_{ij}}\right)^d & \, d = 1, 2, \dots \end{array}\right. $ \\ 
			\rule{0pt}{3.5ex}	& where $\Gamma$ denotes the gamma function. \\
				& \\
			$\operatorname{Binomial}\left(N, \pi_{ij} \right)$ & $ \operatorname{Pr}\left( D_{ij} = d\right) = \binom{N}{d} \pi_{ij}^d \left(1 - \pi_{ij}\right)^{N-d},\qquad \qquad \qquad \qquad \qquad \enspace d = 0, 1, \dots \, N $ \\ 
				& \\
			$\operatorname{Beta-Bin}\left(N, \pi_{ij}, \phi_{ij} \right)$ & $ \operatorname{Pr}\left( D_{ij} = d\right) = \binom{N}{d} {\pi_{ij}^*}^d \left(1 - \pi_{ij}^*\right)^{N-d}, \qquad \qquad \qquad \qquad \qquad d = 0, 1, \dots \, N $ \\ 
			
			\rule{0pt}{3.5ex}	& where $\pi_{ij}^* \sim \operatorname{Beta}\left( \frac{\pi_{ij}}{\phi_{ij}} , \frac{1 - \pi_{ij}}{\phi_{ij}}\right) $\\ \hline
			
		\end{tabular}
		\label{tab2}
	\end{adjustwidth}
\end{table}
\end{appendices}

\end{document}